# Unlocking the Potential of Small Satellites: TheMIS's Active Cooling Technology on the SpIRIT Mission

Miguel Ortiz del Castillo, Clint Therakam, Jack McRobbie, Andrew Woods, Robert Mearns, Simon Barraclough, Stephen Catsamas, Mika Ohkawa, Jonathan Morgan, Airlie Chapman, Michele Trenti
The University of Melbourne, Melbourne Space Laboratory
Grattan Street, Parkville, Victoria, 3010, Melbourne, Australia
miguel.ortizdelcastillo@unimelb.edu.au

**ABSTRACT**

The Thermal Management Integrated System (TheMIS) is a key element of the Australia-Italy Space Industry Responsive Intelligent Thermal (SpIRIT) mission, launched in a 510km Polar Sun-Synchronous orbit in December 2023. SpIRIT is a 6U CubeSat led by The University of Melbourne in cooperation with the Italian Space Agency, with support from the Australian Space Agency and with contributions from Australian space industry and international research organizations. The TheMIS subsystem actively cools and controls the temperature of sensitive instruments, increasing the potential range of payloads supported on small spacecraft systems. TheMIS core functionality is based on a commercial Stirling Cycle Cryocooler in-principle capable of reaching cold-tip temperatures below T=100K. The cooler is operated by customized control electronics and is connected to deployable radiators through pyrolytic graphite sheet thermal straps, all developed by the University of Melbourne. Until now, this level of thermal control has been relatively uncommon in nanosatellites. On SpIRIT, TheMIS aims to validate the design and performance by controlling the thermal environment of SpIRIT's HERMES payload, an X-ray instrument provided by the Italian Space Agency which has a noise background strongly sensitive to temperature. Beyond SpIRIT, TheMIS has the potential to support a broad range of applications, including holding infrared focal plane arrays at cryogenic temperatures, and increasing resilience of electronics to space weather. This paper provides an overview of TheMIS's design, implementation, and operational performance, detailing the commissioning phase and the early results obtained from its operations in orbit, with comparison to the thermal model developed during the mission environmental testing campaign. Finally, the paper discusses ongoing challenges for thermal management of payloads in small satellite systems and potential future strategies for continuous improvement.

## INTRODUCTION

Thermal management is a key element of satellites for remote sensing, and solutions ranging from passive cooling to cryostats have been implemented on large satellites [1]. The larger the surface area, mass, volume, and available power of a satellite, the easier it is to cool its internal subsystems. Very few small satellites have demonstrated advanced thermal control [2,3]. Therefore, the use of low-noise infrared sensors for Earth observations or astronomy is limited on nanosatellites due to the lack of cryogenic thermal control [4], which can maintain a detector at cryogenic temperatures with minimal temperature variations [5].

Given the challenges of designing and developing a nanosatellite with advanced infrared remote sensing capabilities [6], we addressed an intermediate technology demonstration step. Our aim is to build heritage in thermal control under less stringent operational requirements while adopting design choices that can be utilized for future, more advanced applications with minimal modifications. In this context, the TheMIS (Thermal Management Integrated System) payload, developed by the Melbourne Space Laboratory at the University of Melbourne (UoM), is designed for the SpIRIT nanosatellite mission [7] (see Figure 1).

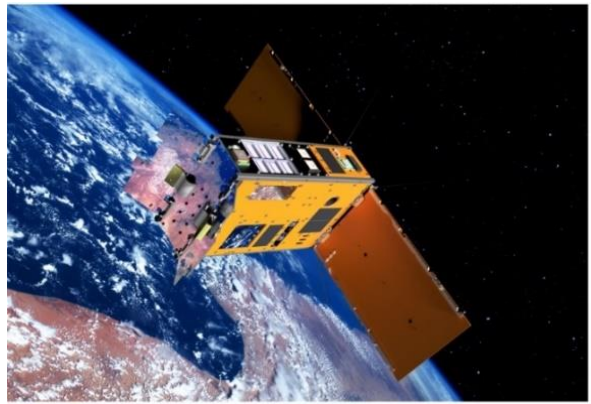

**Figure 1: SpIRIT satellite render in fully deployed configuration – TheMIS radiators and both solar panels deployed.**

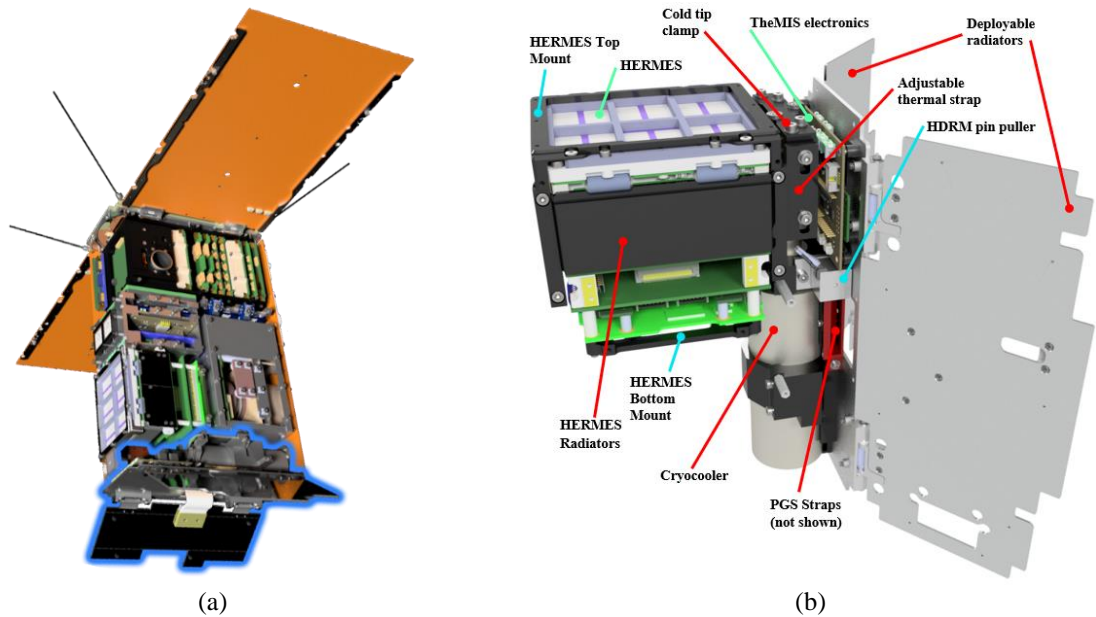

Figure 2. (a) TheMIS payload location at the bottom of the spacecraft highlighted. (b). TheMIS architecture highlighting key attributes including HERMES, cryocooler, Pyrolytic. Graphite Sheet (PGS) straps and deployable radiators.

The SpIRIT mission is the first funded by the Australian Space Agency to carry a foreign agency payload—the HERMES instrument, an X-ray detector provided by the Italian Space Agency [8] [9]. To minimize noise in the detector, HERMES requires a cool and controlled thermal environment—with temperatures between 0°C to −30°C. Within this range, colder sensor temperatures minimise sensor noise. To this end, TheMIS uses a Commercial off-the-shelf (COTS) Stirling cycle cryocooler from as a thermal actuator.

In addition, TheMIS incorporates control electronics developed by the University of Melbourne, pyrolytic graphite sheet thermal straps, and one-way deployable radiators with an adjustable hold-down and release mechanism. These active and passive thermal management technologies enable TheMIS to enhance the performance and data quality of the HERMES sensor, all within a compact form factor.

The remainder of this paper is organized as follows. First, we present the concept design, followed by an introduction to the different building blocks and functionalities. Next, we introduce the payload architecture and its final design. Subsequently, we describe the radiator performance optimization technique, and finally, we present the early results of the payload in-orbit.

## THEMIS CONCEPT

TheMIS is a thermal management module designed for nanosatellites, specifically developed to provide active cooling for satellite-hosted instrumentation. It is designed for applications that require cryogenic temperatures down to temperatures of circa 80K at the cooler's cold tip, although operational parameters such as temperature can vary depending on specific mission requirements, in which case the TheMIS design architecture can be tuned for warmer temperatures (as has been done for SpIRIT). TheMIS is applicable to a diverse range of applications, ranging from infrared remote sensing and high-energy radiation detection, to enhancing the resilience of electrical systems within harsh radiation environments [6].

In the SpIRIT mission, the purpose of TheMIS is to cool the HERMES detector electronics within a 0°C to −30°C range with better than 1°C stability, which is expected to substantially reduce the temperature and radiation-dependent instrumental noise level [10]. This capability could extend the operational effectiveness to higher background regions, such as higher latitudes and possibly the South Atlantic Anomaly (SAA) for future iterations at colder temperatures. Demonstrating consistent functionality in these high-radiation zones of the HERMES detector is one of the primary objectives of the SpIRIT mission.

TheMIS electronics performs the transformation of power from the DC provided by the satellite's platform to the AC required by the cryocooler. This unit not only converts power to drive the cooling device but also supports multi-point temperature measurement. The primary actuator in TheMIS is a COTS tactical cryocooler, operating on a reverse Stirling cycle. The cryocooler extracts heat from the payload detector, particularly the front-end electronics, and rejects it into space via the radiators, ensuring a controlled thermal environment for the payload. The cryocooler is strategically placed within the spacecraft to optimize heat rejection through the deployable radiators. Additionally, TheMIS includes a software control loop to consistently maintain thermal equilibrium at set points by adjusting the power supplied to the cryocooler. Figure 2(a) shows the position of TheMIS within the SpIRIT nanosatellite's mechanical structure, Figure 2(b) presents a detailed view of the payload components, including the cryocooler, the deployable radiator, and the electronics that manage power and thermal regulation.

**PAYLOAD ARCHITECTURE**

TheMIS comprises of three primary components: the cooler electronics, deployable radiators, and the cryocooler equipped with thermal straps. Figure 3 and Figure 4 illustrate key mechanical (blue), thermal (red) and electrical (green) features of the Flight Model (FM) TheMIS subsystem. The total mass of FM TheMIS was ~1800 g (including the 1000 g cryocooler), and FM HERMES was ~1700 g.

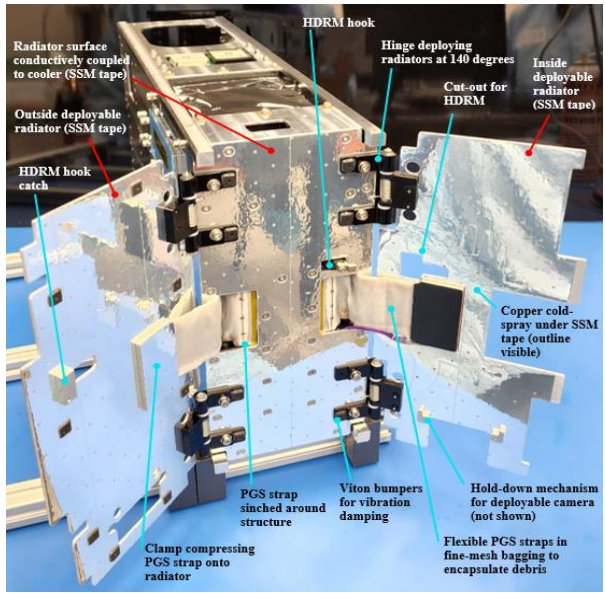

**Figure 3. Back of FM TheMIS (deployed). Outside radiator (left), MZ closure panel/radiator (middle), inside radiator (right). Mechanical features (blue), and thermal features (red).**

*Radiators*

The deployable radiators, held at a 140-degree angle by a torsional spring and a hinge stop, are shown in their extended configuration in Figure 3. All three radiator surfaces—the middle radiator (MZ-closure panel) and the two deployable radiators—are visible and coated with Second Surface Mirror (SSM) tape. This tape, characterized by high infrared emissivity (>0.75) and low visible absorptivity (<0.1), is silver-backed and Teflon-coated. Only the visible side of the two deployable aluminium radiators are coated in SSM tape, with specific areas fine-tuned based on thermal simulation outcomes. The non-space facing sides (back sides of deployable radiators not seen in Figure 3) were left with a bare metal finish, based on the predictions of the preliminary thermal simulations. This design approach, including the oversizing of the total radiator area, was chosen to allow for potential future increases in radiator heat rejection capability. This adaptability could be particularly useful following the SpIRIT FM thermal balance correlation, as thermal balance testing will reveal the true physical thermal performance of TheMIS and SpIRIT as a whole.

The utilisation of SSM tape, coupled with the substantial radiative view factor to space, enables efficient heat rejection from the cooler's rear in the deployed configuration. In contrast, in the stowed configuration, the right radiator plate (inside radiator) folds inward while the left plate (outside radiator) closes over it. A latch linked with the Hold-Down and Release Mechanism (HDRM) secures the stowed assembly, which fits within the narrow space (<10 mm) behind the spacecraft rails.

During the Launch and Early Operations Phase (LEOP) and the initial characterization of payload operations, the HDRM functions as a single-fire heat switch, keeping the panels stowed and exposing only their external low-emissivity (bare aluminium) side to space. Once the spacecraft and payloads are fully commissioned and ready for the main mission objectives—including cooling HERMES—the radiators will be deployed for the remainder of the mission. This deployment strategy ensures sufficient time for the commissioning of the Attitude Determination and Control System (ADCS) and allows any parasitic heat from the spacecraft payloads to warm the back end (TheMIS side), thereby mitigating the risk of overcooling SpIRIT. This approach has successfully kept TheMIS within operational and non-operational limits during the in-orbit commissioning phase of SpIRIT.

Pyrolytic graphite, noted for its high thermal conductivity and exceptional qualities as a thermal interface material, offers significant mass and volume advantages over traditional strap materials such as aluminium and copper. Specifically, the selected pyrolytic graphite sheet (PGS) has a thermal conductivity of 700 W/(m·K) for 100 μm straps [11]. We integrated these PGS straps beneath the mechanical structure of the cooler, utilizing them both as the strap material and as a thermal interface. The straps thereby provide a conductive path between the cooler body and deployable radiators, as detailed in Figure 3. PGS is flaky and is electrically conductive, therefore PGS thermal straps must be bagged to prevent debris interfering with spacecraft systems. As such, vacuum-compatible fine mesh encases the PGS straps. The design of the straps and their cutouts allow for the flexibility necessary to safely transition between configurations with radiators stowed and deployed.

Figure 4 presents the front view of the TheMIS payload, displaying the mechanical mounts and clamps for the cryocooler, the HDRM pin puller (supplied by DcubeD), and the TheMIS electronics. The cooler compressor and the cooler cold head are mounted on separate mechanical seats, both thermally coupled to the structure. The transfer line connecting the compressor and cold head is supported by small custom blocks (as they are thin walled and fragile). The cooler cold tip connects to the HERMES thermal strap assembly (not shown).

Upon activation, the pin puller draws a rod through a channel within the cooler cold head mount (beneath the cooler cold head). This rod is connected to a hook that can pivot, transforming the linear motion of the pin retraction into the rotational movement of the HDRM hook. This action releases the hook catch, allowing for the deployment of both radiator panels and the inspection camera assembly (not shown). As seen in Figure 3, the inner radiator (right) features substantial cutouts to accommodate the hook release and folding over the pyrolytic graphite straps. The reduced surface area and constrictions near the compressed straps result in diminished radiator performance. To address this, two thin layers (150 μm each) of cold-sprayed copper is applied to either side of the inner radiator, enhancing heat distribution while minimizing mass. The application of sprayed copper is strategically designed to optimize thermal performance (minimise average temperature at the region where the cooler-radiator strap meets the radiator) within a mass and volume constraint.

*Cryocooler*

A cryocooler is an electro-mechanical device that utilizes a compressor and a working fluid—helium gas in this instance—to provide cryogenic cooling. The cryogenic cooler from Thales Cryogenics is a compact closed-loop, Stirling-cycle cooler comprising a compressor module and a cold finger, which are interconnected by a transfer pipe. The compressor's pistons, driven by integrated linear electric motors, are gas-coupled to a free displacer in the cold finger.

For the SpIRIT mission, preliminary studies evaluated two different coolers from Thales Cryogenics: the UP8197 and the LSF9987. Initially, the UP8197 was chosen as the baseline cooler due to its smaller size, which is advantageous for volume-constrained CubeSat applications. However, the UP8197 was unavailable at the time of cooler selection. From an electronic compatibility perspective, both cooler options are similar, with TheMIS electronics engineered to meet the drive requirements of either model. The primary distinction between the two designs lies in their drive (and consequently, resonant) frequencies. In both scenarios, the electronics are capable of driving dual linear electric motors and managing the cold-tip temperature via a closed-loop control system.

However, the SpIRIT mission opted for a slight modification to the LSF9987, selecting the LSF9997 model. This model is an integrated dewar cooler assembly (IDCA) cooler that employs the same miniature moving-magnet flexure-bearing compressor as the LSF9987, which is a slip-on cooler. The primary difference is the cold tip, with the LSF9997's being slightly smaller in length, enabling it to fit in the 2U volume designated for SpIRIT. This cryogenic cooler is designed to achieve and maintain operating temperatures typically ranging from 60 to 115 K, ideal for cooling sensors and electronics. In the case of SpIRIT, this temperature is significantly lower than required. It is important to note that the TheMIS design for SpIRIT does not require, and as a result is not engineered, to cool HERMES to temperatures as cold as theoretically possible.

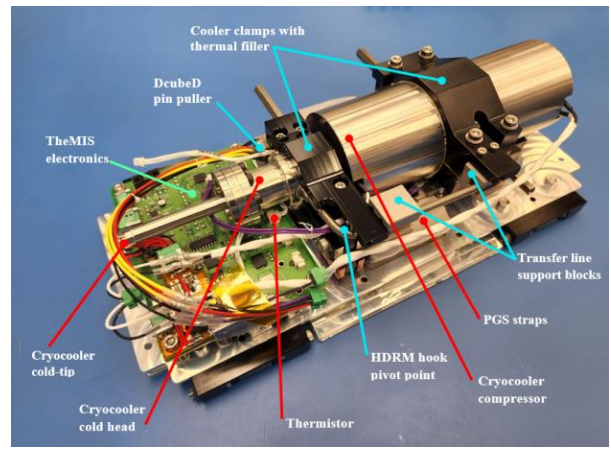

**Figure 4. Front of FM TheMIS (stowed). Mechanical features (blue), thermal features (red), electronics (green).**

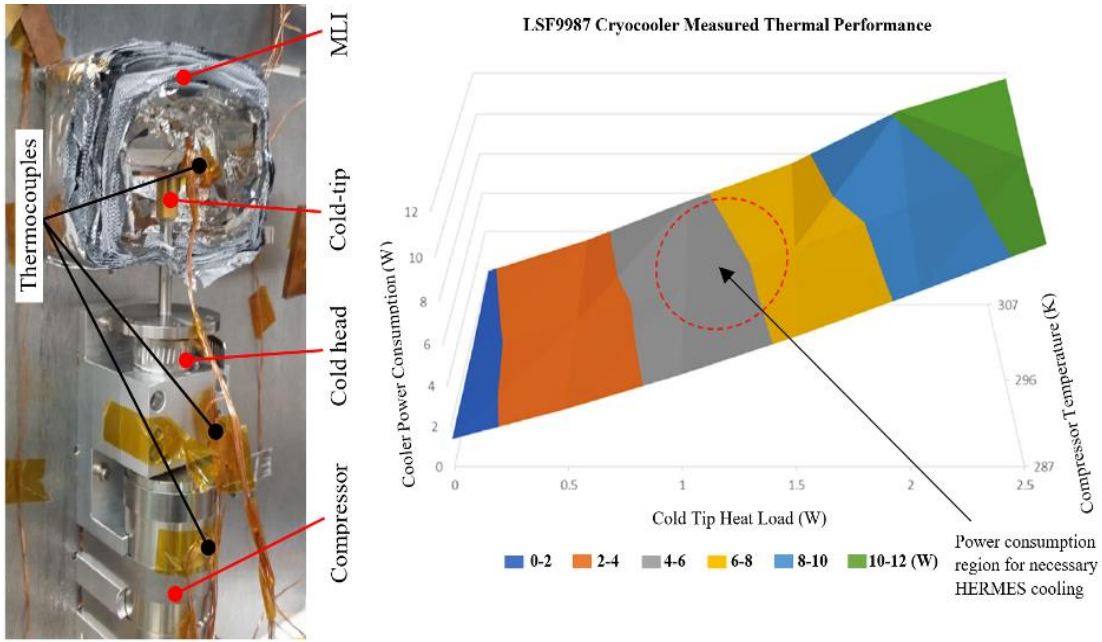

**Figure 5. LSF9987 Cryocooler under test (left) and measured performance for SpIRIT operation (right).**

While Thales Cryogenics provides control electronics for these coolers, those units are not suited for space applications. Consequently, the University of Melbourne has developed bespoke control electronics in-house. Among the various operating modes available with the University of Melbourne control electronics, a key feature for the SpIRIT mission is "On-Demand Refrigeration" The detector temperature is controlled by closed loop control using a Resistance Temperature Detector (RTD) thermally coupled to the cold strap to measure the target temperature. This setup ensures that any excess heat generated by the payload is efficiently managed "on demand." The design of the control loop takes into account the thermal transfer characteristics (time constants) of the cooler-detector combination, which are critical for tuning the controller parameters.

To date, the LSF9987 variant of the cooler has been tested at the University of Melbourne. Characterization testing of the LSF9987, conducted under vacuum (Figure 5), has indicated that the cryocooler's power consumption ranges between 6 and 8 W to dissipate the heat generated by the HERMES detector (0.19 W) and the parasitic heat from the spacecraft (1 W to 1.5 W). Tests on the LSF9997 have shown comparable performance, affirming the relevance of the LSF9987's characterization data. The integration of SpIRIT's LSF9997 cooler into the SpIRIT structure has been successful, demonstrating that a cryocooler and its electronics can be accommodated within a 2U volume.

A significant challenge with these COTS tactical coolers involves managing the pressure of the working fluid. Discussions with several Launch Service Providers (LSPs), including SpIRIT's launcher SpaceX, have revealed that pressure vessels operating above 250 psi (17.4 bar) require qualification testing to verify that the mechanical structure can safely handle 1.5 times the Maximum Expected Operating Pressure (MEOP). Given that the design of these coolers necessitates helium pressures surpassing this limit, qualification tests were essential. Thales Cryogenics has successfully conducted these tests for the SpIRIT mission, confirming that the coolers meet the necessary LSP requirements for pressure compliance.

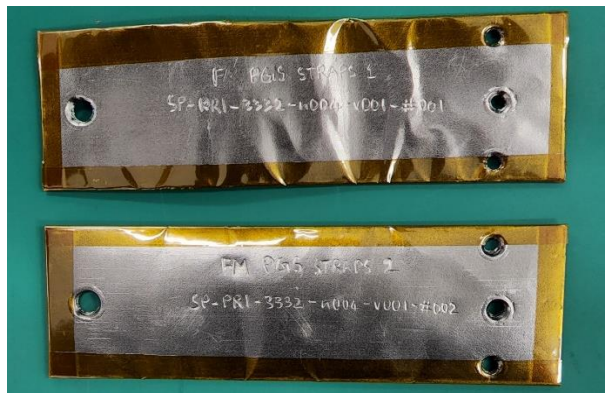

**Figure 6. SpIRIT flight PGS thermal strap set. Each half strap is for one radiator.**

*Pyrolytic Graphite Sheet Thermal Strap*

Flexible thermal straps are essential for transferring heat from the cryocooler compressor to the deployable radiators in the TheMIS system. Initially, copper straps were considered due to their robustness, but they presented challenges such as lower thermal performance, interface space issues, and higher mass. Consequently, pyrolytic graphite sheet (PGS) straps were selected for their superior flexibility, high material and interface conductivity, and low mass.

At the University of Melbourne, experiments have been conducted to optimize the composition of the PGS straps, experimenting with various sheet thicknesses and layer counts. These straps are designed to dissipate a total of 10 W, with each strap handling 5 W of heat. In-house analyses led to extrapolations for optimal conduction properties. PGS increases in thermal conductivity with reducing thickness—though thinner sheets increase in density. For instance, 100 μm and 70 μm straps have a thermal conductivity of 700 W/(m·K) and 1000 W/(m·K) respectively [11]. Initially thinner sheets up to 25 μm were considered for their higher thermal conductivity but assembly attempts demonstrated that the thin sheets were prone to tearing and damage. Consequently, the Flight Model (FM) employs 10 layers of the more robust 100 μm thick PGS, resulting in a total mass of 14 g per strap set—significantly less dense than copper (copper is 10x the density of 100 μm PGS) but more durable than the thinner PGS variant.

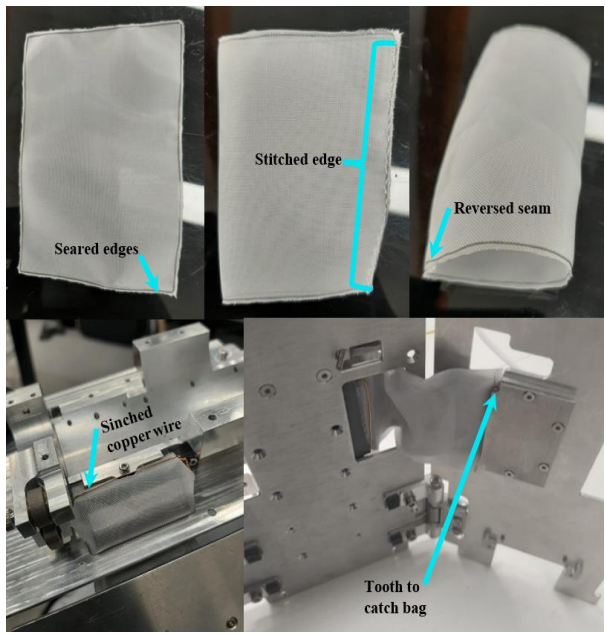

**Figure 7. EM example PGS strap processing and deployment.**

Despite their advantages, PGS straps are susceptible to flaking and debris generation, which can be problematic for optical payloads and electronics. To mitigate this, COTS polypropylene mesh was used to encase the straps. The assembly includes custom 3D-printed lipped mounts and a stepped feature on the radiator compression plate, which allows the bag to be tightly sealed at both ends. During the cutting and handling of the bagging material, the edges were found to be prone to fraying. To address this, a method was developed using the radiative heat from a hot soldering iron to melt the edges in a controlled manner (see Figure 7). The bags were then stitched and sealed along the seams with epoxy at both ends to prevent any debris escape.

*HERMES Thermal Strap*

To thermally couple the HERMES detector to the cryocooler cold-tip, an aluminium strap was designed to interface with the cold-tip and the HERMES mechanical structure. The region of interest of the HERMES detector that requires cooler (for noise reduction and radiation resilience) is the front-end electronics The two HERMES front-end electronics PCBs, located directly behind the HERMES radiators (see Figure 8), have been thermally coupled to these radiators using thermal filler. Maximising thermal coupling between the HERMES radiators and these PCBs is achieved by applying even compression across the interface with four bolt points on each radiator, using washers as shims to accommodate for manufacturing tolerances.

Figure 8 depicts the configuration of the HERMES thermal interfaces. Initially, a flexible thermal strap consisting of interface blocks and copper sheets/braids was considered for the conductive link between HERMES and the cold-tip. Additionally, the placement of mass at the cold-tip of the cooler creates a cantilevered beam effect, which could lead to excessive vibration during launch, potentially damaging the thin-walled cold finger tube (0.1 mm). Consequently, it was determined that integrating the mechanical support and thermal interface into a single, simplified system would reduce complexity and risk. Stresses applied to the cold finger tube can easily damage it. Given the manufacturing tolerances of the multiple mating surfaces and the challenges associated with precisely bending the cooler's transfer line, the centroid location of the cold tip could vary significantly (shifted up to 6 mm x 2 mm x 2mm in respective axes). Therefore, the thermal link and clamp point are designed with a high degree of adjustability, not relying on a single machined link. This adaptability is facilitated by multiple slotted holes as shown in Figure 2 and Figure 8. To improve the thermal interface between the link components, a thin indium thermal filler is placed between the components, also used between the cold-tip and its clamp.

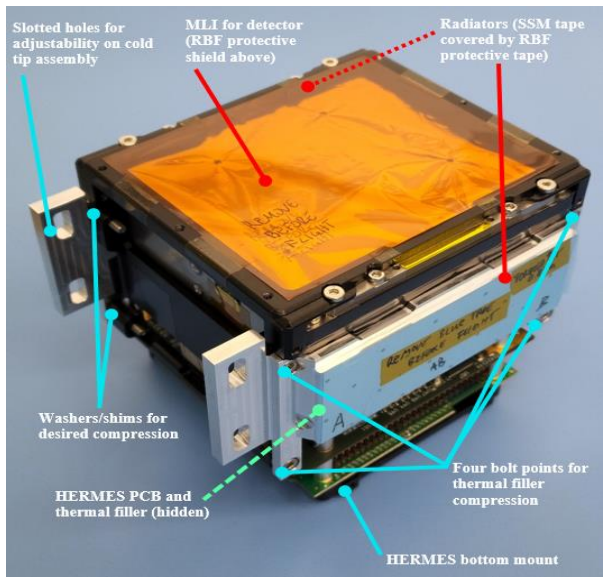

**Figure 8. FM HERMES assembly showing only radiators and cold tip assembly interface. Mechanical features (blue), thermal features (red), electronics (green).**

Bare aluminium, static radiators on the exterior of the HERMES payload were covered with SSM tape and are primarily space facing, which handles the majority of heat rejection from HERMES. Thermal simulations predict these radiators can passively cool HERMES to around $-18°C \pm 1°C$. The cryocooler then precisely controls the cooling of the detector, maintaining temperature stability instead of the spacecraft's cyclical transient temperatures driving the detector temperature. This setup offers capacity for passive cooling while the TheMIS payload is non-operational. To ensure optimal heat rejection, the area covered by SSM tape was adjusted post-system thermal testing. This tuning capability provides a backup in the thermal design against overcooling. Analysis and correlation of the spacecraft thermal model after Thermal Balance (TBAL) demonstrated the HERMES radiator sizing was appropriate for flight and therefore was left as is.

SSM tape was selected for the SpIRIT radiators. While alternative coatings were considered, they proved impractical. Thermal paints are difficult to source in Australia, and options such as Z306 paint have a poor shelf life. Although paints and tapes such as Z306, Kapton tape, and black Kapton may offer higher infrared emissivity, they also have much higher solar absorptivity (5x to 10x SSM tape). Mirror tiles, while very high in IR emissivity and very low in visible absorptivity, were prohibitively expensive and impractical for SpIRIT. Given these factors and the unique shapes and cutouts of the radiators, SSM tape remained the preferred choice.

*Hold Down and Release Mechanism*

Traditional nanosatellite burnwire release mechanisms, which use resistors operating beyond their optimal thermal range, often fail to achieve the necessary temperatures for deployment. This method has resulted in unreliable deployment across several missions [12]. Given the important role of the TheMIS radiator deployment in dissipating heat from the cryocooler and providing further passive cooling for HERMES (in addition to the HERMES passive radiators), a highly reliable deployment mechanism is critical for TheMIS. Consequently, a compact, high-reliability COTS pin puller (TRL 9) from DcubeD was selected as the release mechanism for the deployable radiators. This pin puller activates by utilising a heater circuit to enable a pin retraction, which is linked to an adjustable lever mechanism designed by the University of Melbourne, thereby enabling the release of the radiators. Deployment tests conducted in ambient conditions using this pin puller mechanism have consistently succeeded, demonstrating its reliability with 40 successful activations (33 EM, 7 FM). The EM spacecraft (which was mechanically similar) underwent vibration loads of 70% GEVS Qualification. These loads are far greater than the loads SpIRIT was subject to on the way to orbit onboard SpaceX's Falcon 9. The post vibration test deployment of the TheMIS radiators was successful, affirming the design's maturity and increasing confidence in successful in-orbit deployment.

Figure 9 illustrates the HDRM system schematic, outlining two adjustable elements crafted for ease of assembly. The hook mechanism is designed in two parts: the main lever and an adjustable hook that can be detached, allowing the entire radiator assembly to be externally stowed. A custom Viton rubber bumper is placed between the latch and hook to prevent metal-on-metal contact caused by the hook's momentum post-deployment. The adjustability of this system is vital to compensate for cumulative machining tolerances, ensuring the hook does not grip the radiator too tightly—which could overwhelm the pin puller's capacity and cause deployment failure—or too loosely, which might lead to rattling and potential damage to the radiator or HDRM assembly.

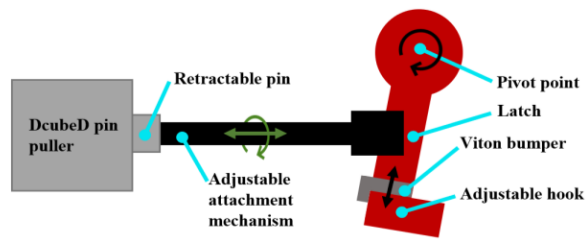

**Figure 9. Hold-down and release mechanism schematic.**

# RADIATOR PERFORMANCE OPTIMIZATION

## Numerical Technique Optimization Theory

Optimization of radiators has been previously investigated through numerical techniques as the complexity of such designs simply cannot be achieved easily by hand. Optimizing slightly complex radiator geometries, or the presence of non-symmetrical heat loads into the radiator, will lead to non-trivial solutions. Shape optimization of space radiators has been explored via genetic algorithm and in particular for the SpIRIT radiator design, researchers investigate the optimal shape and distribution of highly conductive graphite on a radiator via particle swarm optimization [13] [14]. Optimization techniques including simulated annealing were applied to study novel space radiator designs [15]. Here we seek to build upon these studies by applying optimization methods to design higher performance radiators to fly on the SpIRIT nanosatellite.

To perform optimization on the SpIRIT radiators, a numerical model was first needed to evaluate the performance of each design. Analysis of the performance was conducted for steady-state operation with the boundary conditions of a given heat flow applied to defined regions of the radiator plate. The radiator was modelled as an isolated body in free space, with radiation and conduction being the sole methods of heat transfer. While the radiators would be subject to environmental flux factors (Earth IR flux, Earth albedo flux and solar flux), the focus of the design was on steady state operating conditions, ignoring transient environmental perturbations. The assumption was also made that the plate is thin and thus can be analysed as a 2D problem. Under these assumptions, the heat flow can be modelled by the steady-state heat equation:

$$-\nabla \cdot (\kappa \nabla T) = \dot{q}_{app} + \dot{q}_{rad} \quad (1)$$

Where T is the temperature of the plate, $\kappa$ is the thermal conductivity per square, $\dot{q}_{app}$ the applied heat flux density, $\dot{q}_{rad}$ the radiative flux density. The radiation from the plate, $\dot{q}_{rad}$ is given by the Stefan-Boltzmann law as:

$$\dot{q}_{rad} = 2\sigma\varepsilon(T^4 - T_b^4) \approx 2\sigma\varepsilon T^4 \quad (2)$$

Where the parameter $\sigma$ is the Stefan-Boltzmann constant and $\varepsilon$ is the emissivity of the radiator plate. The constant of 2 arises from the plate radiating from both sides, the background temperature $T_b$ may be ignored as the radiator will be operating at temperatures near 300 K, significantly greater than the 3 K background temperature. Thus, solving the system:

$$-\nabla \cdot (\kappa \nabla T) - 2\sigma\varepsilon T^4 = \dot{q}_{app} \quad (3)$$

We solved for the temperature of the plate. To do this numerically a finite difference scheme was adopted and solved for the temperature $T^i$ on each grid node i via:

$$2\sigma\varepsilon(T^i)^4 + \sum_{j \in N} \kappa_{ij}(T^i - T^j) = \dot{q}_{app}^i \quad (4)$$

Where N is the von Neumann neighborhood, $\kappa_{ij}$ is the conductivity matrix, whose coefficients depend on the material of the nodes i and j. The solver used to determine the temperature on the plate was verified on various sample problems with industry-standard thermal modelling tools including ESATAN, and NASTRAN.

For high radiator thermal performance, the optimal distribution of a copper pattern on the aluminium plate was sought after. The copper layer is more effective at distributing the heat from heat input at the thermal strap interface as copper has a higher thermal conductivity than aluminium. Specifically, the cost function of the maximum temperature on the plate is chosen. This cost function is chosen as the lower the temperature at the thermal interface, the lower the thermal load on the cryocooler.

In this study, we consider binary optimization for the placement of copper on an aluminium plate. The model assigns each grid cell as either empty space, aluminium, or a combination of copper and aluminium. The optimizer takes material properties and three image files as inputs. These image files specify: (a) the shape of the aluminium radiator and the initial distribution of copper, (b) the applied heat flux density at each grid cell, and (c) the keep-out regions where copper should not be placed by the optimizer.

The optimization is performed via a modified form of simulated annealing [16]. Simulated annealing seeks to mimic the physical process by which a slowly cooling system will gradually converge into its lowest energy state [17]. In its implementation from an initial state, a small random perturbation is introduced. If the difference in the cost function is negative or slightly positive, the perturbation is accepted; otherwise, it is rejected when the annealing temperature (T) exceeds a small threshold his temperature is gradually decreased until a final state is achieved [18]. The ability of simulated annealing to accept states with a higher cost function facilitates the exploration of the space, allowing the algorithm to escape local minima [19]. In this context, a deterministic threshold update rule is employed for accepting new states instead of a stochastic one as previous studies indicates that a deterministic threshold can perform similarly to stochastic rules [20].

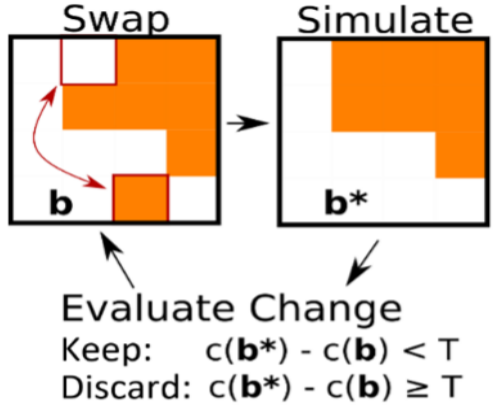

**Figure 10. Optimization algorithm loop used in the radiator design.**

*Numerical Technique Implementation*

Starting with an initial copper distribution $b$, a grid cell of copper and a grid cell of aluminium, both selected from the copper-aluminium boundary, are swapped to create a new distribution $b^*$. Restricting swaps to boundary grid cells was found to increase the optimizer's convergence rate. The cost function, $c$, which measures the maximum temperature on the plate, is evaluated for both distributions. If the change in cost, $c(b^*) - c(b)$ is less than the current annealing temperature $T$, the new distribution $b^*$ is accepted; otherwise, the original distribution $b$ is retained. This grid-cell swapping process is repeated iteratively. Traditionally, the process continues until only marginal improvements in the cost function are achieved. However, for these radiator designs, we instead set a fixed number of iterations (60,000) for all runs. This choice was based on trial runs showing diminishing returns around this iteration count, with improvements less than $1.0 \times 10^{-5}$ K. The annealing temperature $T$ is gradually decreased to zero throughout the optimization. An overview of the optimization loop is provided in Figure 10. Future work will focus on developing a more efficient method for determining the simulation endpoint, as the current fixed iteration approach, while functional, is not optimal and requires improvement.

The inner radiator plate, featuring constrictions, was chosen for optimization using our optimizer. Initially, the side with the greater allowable volume (clamp side) was selected for analysis. To evaluate the optimizer's performance, 50 human-generated copper distributions were tested to determine their maximum temperature. These patterns, based on intuitive reasoning to reduce the maximum plate temperature, were tested at copper thicknesses of 0.8, 1.6, and 3.2 mm, resulting in a total of 150 radiator designs. Each design was then processed by the optimizer to generate potentially more optimal configurations with the same copper thickness and mass constraints. As shown in Figure 11, the optimizer was able to further minimize the cost function compared to human designs. On average, a decrease of 1.0 K in the maximum temperature was observed for designs with the same mass of copper, representing a significant portion of the maximum temperature reduction seen in uniform copper designs. The smallest observed decrease between pre- and post-optimization was 0.3 K, indicating the optimizer's ability to outperform even the best human designs. The total CPU time for all 150 optimization runs was 6 hours, using an i5-8265U CPU with a turbo clock speed of 3.9 GHz.

The optimization aids in the design of the heat spreader and provides a greater understanding of the problem space and design decisions. For instance, the dense sampling of different copper masses produces quantitative data on the diminishing returns achieved from greater amounts of copper and the data on the thickness of copper indicates relative insensitivity to thickness except as the copper begins to cover a substantial portion of the aluminium radiator plate.

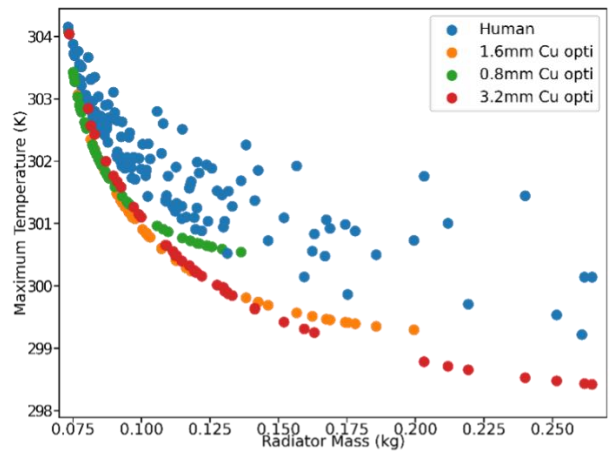

**Figure 11. Maximum temperature vs radiator mass for various patterns of copper on the radiator panel.**

*Optimized Radiator Design*

Based on further data collection and analysis, it was found that applying copper on the back side of the radiator (opposite side of the clamp) was far more effective and mass-efficient than the front. Applying copper on the front side meant that copper could not be applied in the clamping region, which was the best region to spread heat. This finding in itself was surprising considering that the front side could allow for 3.2 mm thick copper if desired. On this side of the radiator, initially based off the CAD, there was a 1.0 mm gap between the two radiators in the stowed configuration. Thus, 0.8 mm maximum copper thickness was initially allocated to work with Despite the limited thickness allowed, the back-side copper spray outperformed the front-side options; quick and easy iterations with this tool allowed for such critical findings. An 80-20 rule was applied to determine an optimal maximum temperature reduction vs mass trade-off. As can be seen in Figure 11, there are diminishing returns for increasing masses. Hence, the copper mass at which 80% of the total possible maximum temperature reduction was selected. It was determined that a copper thickness of 0.8 mm and a total plate mass of 0.105 kg (copper mass of 0.031 kg) was the best compromise between performance and mass. A picture of the final radiator design is displayed in Figure 12. This radiator provided a 10.5 K decrease in the maximum temperature (compared to the bare aluminium plate), far superseding even the 3.2 mm thick copper designs for the front side. Applying SSM tape on this plate (including on top of the copper), provides further tuning of the radiator thermal performance.

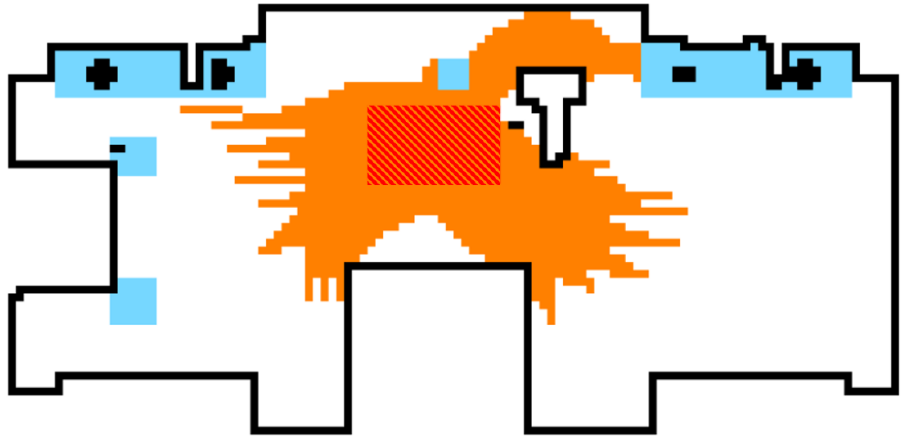

**Figure 12. Optimized radiator: 0.8mm Cu thickness, 0.105 kg total mass. White: Aluminium, Orange: Copper, Red: heat input zone, Blue: keep out zone. Note that the red and orange hash indicates a shared copper and heat input zone.**

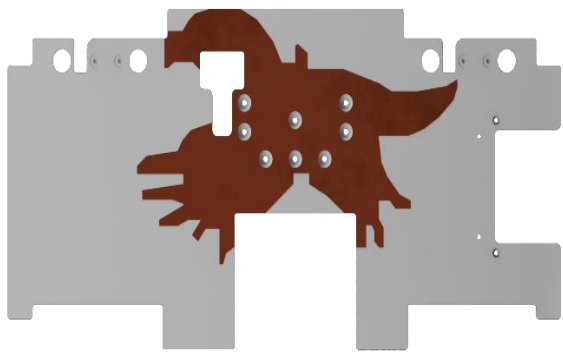

**Figure 13. Simplified inner deployable radiator. Silver: Aluminium, Red: Copper.**

While the current iteration of the optimizer has been successful in optimizing radiator designs and informing design choices, it has several limitations. It only considers the binary presence of copper, rather than allowing for a continuous thickness distribution. Additionally, the optimizer operates without incorporating the underlying physics of the problem, limiting its ability to find truly optimal solutions. Furthermore, the model is isolated and does not account for the thermal environment and dynamics in which the radiator will be deployed. To address these limitations, future work could involve applying topological optimization methods and developing a more detailed thermal environment model, thereby improving the optimizer's accuracy for real-world applications.

## MANUFACTURING OF THE COLD SPRAY COPPER-ALUMINIUM RADIATOR

With the limited space available for the radiators, and to leverage local capabilities, cold spray additive manufacturing was selected as the preferred method to generate a bimetallic (aluminium and copper) radiator. Cold spray is a coating deposition process that involves accelerating metal powders, via a gas, to plastically deform and adhere the powder to a substrate. While the numerical model provided the optimal copper pattern for the described constraints, the pattern displayed in Figure 12 required simplifications in the regions of very thin members—in order to avoid sections breaking off and to ensure good cold spray adhesion with the aluminium substrate (see Figure 13).

Several test panels were coated to characterise the temperature of the copper powders, travel rate of the arm, layer thickness and stencil thickness (see Figure 14 and Figure 15 for an example of the test spray setup).

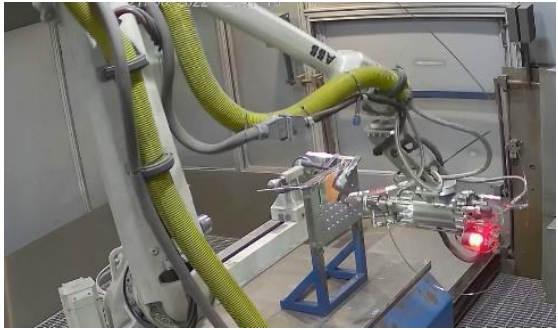

**Figure 14. Titomic cold spray arm coating SpIRIT radiator with copper.**

Early test sprays of 0.8 mm on one side of the radiators demonstrated an unexpected and substantial deformation away from the cold sprayed surface. The radiator deformed with a maximum arc height of 5 mm from the tips, which was greater than 6 times the copper cold spray thickness (see Figure 16).

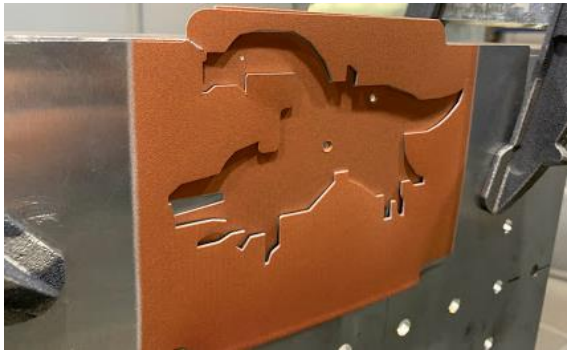

**Figure 15. Preliminary cold spray copper on stencil and radiator. Vice grips clamping specimens.**

Further investigation and research showed that plastic deformation occurs due to the shot peening effect on the cold spray coupon's side. The high-speed impacts of the cold spray copper generate a compressive residual stress field [21]. The distribution of compressive and tensile residual stresses along the impact direction causes the coupon to be convex shaped [21]. This deformation process is often shown on a thin metal strip called an Almen strip. Figure 17 demonstrates the same 'spring back' effect once the coupon is released after shot peening [22].

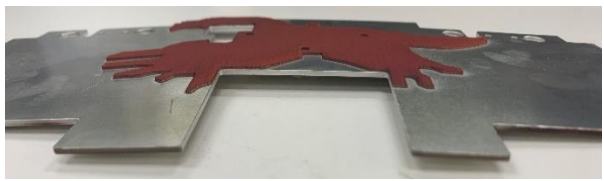

**Figure 16. Sample aluminium radiator deformation (5 mm arc) on with 0.8 mm cold spray copper.**

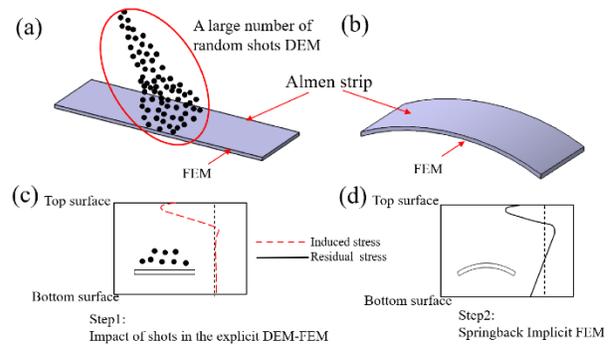

**Figure 17. Improved DEM-FEM method. (a) Impact of shots in the explicit DEM-FEM. (b) Spring back Implicit FEM. (c) Induced stress profile. (d) Residual stress profile. [22]**

As exceeding the volume where the radiators are stowed behind the spacecraft would constitute the spacecraft failing to fit inside the deployment pod, and for assembly reasons, a cold-sprayed panel with this deformation is not useable. Attempts of rolling the radiators back to flat shapes resulted in cracking of the copper and damage to the radiators.

The next attempt was to spray a thinner layer of copper equally on both sides to balance the residual stress. As described in the previous section, based on the 1000 µm gap in the spacecraft CAD, 800 µm thick copper was initially simulated in the numerical model. However, physical assembly of the flight spare radiator on FM SpIRIT demonstrated only a 700 µm gap existed. This was due to not considering the SSM tape thickness (where two layers of SSM tape, middle radiator and inside radiator, is ~300 µm thick). Early test sprays at Titomic also showed that only multiples of 100 µm was possible with the spray set-up. Thus, 200 µm thick copper was selected for each side (total of 400 µm within the 700 µm) to provide assembly margin. This method of balancing the stresses was successful in achieving a flat panel. The coarse cold spray copper however required post-spray polishing. As cold spray generates a rough and uneven surface, the polishing and removal of the top surface resulted in a reduction of total copper thickness. The resultant flight radiators had ~150 µm thick copper on each of the two inner radiator sides. Despite this thickness falling short of the planned 800 µm thickness from the numerical model simulations (with a predicted 10.5 K maximum temperature reduction of the radiator), the total 300 µm of added copper is predicted to cause a 7 K maximum temperature reduction of the radiator. The bare aluminium radiator started at 70 g, with each side of copper adding 7 g (10% of the radiator mass). The final flight radiators had a mass of 84 g. See Figure 18 for the final flight inner radiator.

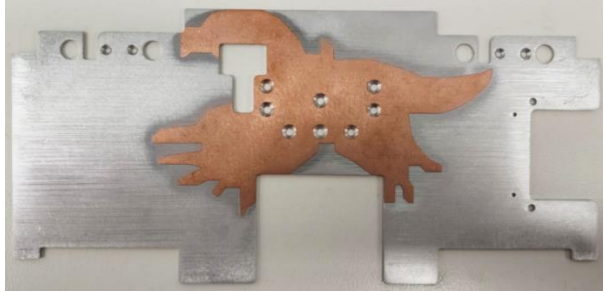

**Figure 18. SpIRIT post-polished flight radiator with ~150 μm copper coated on each side.**

Successful test fits of the copper-coated inner flight radiator within the TheMIS assembly showed that the cold spray process can yield bimetallic flat radiators.

## RADIATOR TUNING POST THERMAL BALANCE TESTING

SpIRIT underwent thermal balance (TBAL) testing as part of the flight environmental test campaign (see Figure 19). As is typical practice, the thermal model of SpIRIT was correlated with the TBAL results to improve confidence in thermal simulation predictions for orbit.

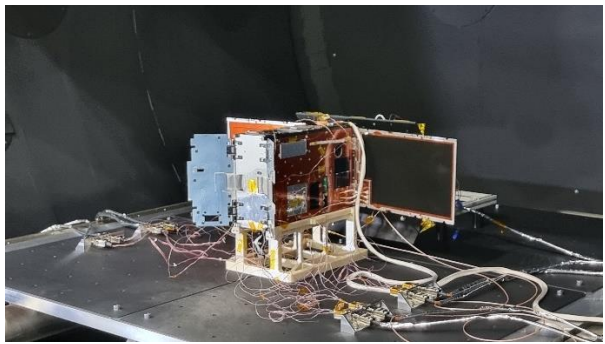

**Figure 19. FM SpIRIT setup for the thermal test campaign in fully deployed configuration. Deployed TheMIS radiators and PGS straps visible. SpIRIT underwent thermal cycling (TCYCLE), thermal vacuum (TVAC), and thermal balance (TBAL) in this configuration. Test conducted in Wombat XL chamber at NSTF, Canberra.**

A key finding from the correlated post-TBAL thermal model was that the conductive couplings at the front end of the spacecraft (opposite end of TheMIS, where the spacecraft bus is located) were higher than anticipated, driving the entire spacecraft colder. The post-TBAL correlated thermal model was re-simulated for orbit, where it was found that the TheMIS electronics and cooler were predicted to reach $-59°C$ as opposed to the $-47°C$ predicted during the preliminary stages of SpIRIT. As the cooler's minimum non-operational temperature is $-55°C$, this new prediction (with no margins) exceeded this limit by 4°C. Furthermore, due to hardware changes to the flight TheMIS electronics (compared to the old preliminary design), the TheMIS electronics' new minimum non-operational temperature limit was updated to $-40°C$. As both the cooler and the TheMIS electronics share the same middle primary radiator (MZ closure panel), when TheMIS is OFF, both these components share similar temperatures. Thus, the TheMIS electronics was also predicted to fail its minimum non-operational temperature (by 19°C).

In order to prevent failure of both the cooler and the TheMIS electronics, radiative tuning of the deployable radiators and/or fixed radiator was necessary. Due to the design of this system, the potential of this surface property tuning was already considered. The final radiator configuration was adjusted to have the middle radiator changed from the high IR emissivity, low visible absorptivity SSM tape to low IR emissivity, low visible absorptivity Vapor Deposited Aluminium (VDA) Polyimide tape. This changed the middle radiator's emissivity from >0.75 to 0.05. See Figure 20 for the impact on the visible and IR properties of the radiators. As expected, the visible properties are similar, while the IR properties of the middle radiator is very low reflecting the body heat of the photographer.

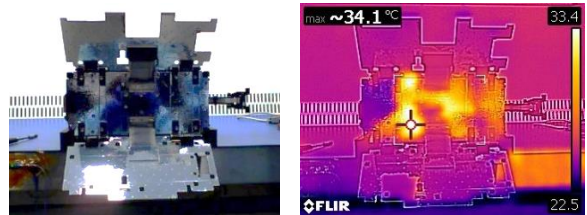

**Figure 20. Post-TBAL tuned FM SpIRIT TheMIS radiators with low IR emissivity aluminium tape applied on the middle radiator, covering the SSM tape (a) Radiators (visible spectrum) (b) Radiators (IR spectrum). Low IR emissivity reflecting body heat of the photographer.**

Re-simulation of the SpIRIT thermal model in orbit, with the VDA tape, demonstrated a simulated increase in predicted minimum non-operating temperature of the TheMIS electronics from an unacceptable $-59°C$ to an acceptable $-37°C$. Further tuning (reduction of IR heat rejection) was possible on the two deployable radiators, whereby doing so would increase the predicted safety margin to the minimum non-operating temperature $(-40°C)$. However, this tuning would then decrease TheMIS' and HERMES' operational performance as a result of hotter passive HERMES temperatures and a reduction in the cooler's potential heat lift. The final TheMIS radiator configuration, post-tuning, only utilised two of the available five surfaces as high emissivity radiators. While this was the correct ratio of high emissivity radiators for SpIRIT, preventing overcooling of TheMIS during low power satellite modes where payloads remained OFF, it demonstrates the heat rejection increase possible for other missions or payloads desiring higher heat rejection. Furthermore, this process demonstrated the criticality of thermal balance testing and correlations to the success of space missions and payloads as it was the correlated post-TBAL model that pointed to the TheMIS cold problem.

**EARLY IN-ORBIT RESULTS**

TheMIS has successfully operated the cryocooler multiple times, using various configurations to showcase its cooling capabilities. These tests have consistently demonstrated the system's ability to effectively manage thermal loads. However, it is important to highlight that the deployable radiators, which are crucial to the overall thermal management system, have not yet been deployed. As a result, the complete cooling capability of the payload will be verified in the upcoming mission phases when these radiators are deployed and tested.

Figure 21 presents the outcomes of the on-orbit cooler commissioning test of 500-second duration. During this test, a reduced gain configuration was used, leading to a notable observation: the temperature on the instrument heat strap decreased by 7 degrees. This result highlights the potential efficiency of the cryocooler even without the deployable radiators in operation. Future tests will provide more data on the system's performance, particularly once the radiators are deployed.

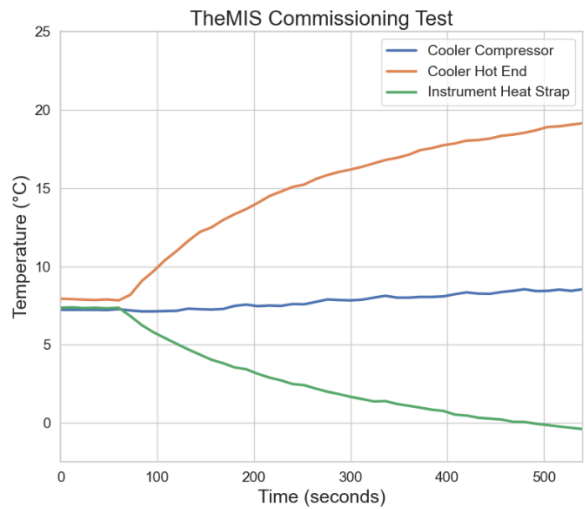

**Figure 21. Themis Commissioning Test during a 500 second run. SpIRIT in radiator stowed configuration during this test.**

**CONCLUSIONS**

This paper has demonstrated the advanced passive and active thermal engineering solutions possible for CubeSats such as SpIRIT. The UoM in-house developed TheMIS payload encapsulates electrical, mechanical, and thermal design features in a compact form factor, utilizing the active cooling capabilities of a COTS cryocooler. The deployable radiator design and pin puller mechanism provide a robust means for deploying radiators on nanosatellites. In particular, the mechanical design of TheMIS has a strong focus on adjustability and ease of assembly, which makes this design suitable for a variety of satellites that intend to cool payloads and instruments—not limited to HERMES. A primary focus of the University of Melbourne's Melbourne Space Laboratory is the development of advanced thermal systems, and the designs outlined in this paper demonstrate this capability. Radiator design with numerical techniques has highlighted some of the more advanced capabilities being developed in the laboratory. The promising early in-orbit results of SpIRIT, launched in December 2023, are the first step towards a complete verification of the system, acting as a precursor for further thermal advances in nanosatellites.

*Acknowledgments*

The authors would like to acknowledge the Australian Space Agency International Space Investment - Expand Capability grant ISIEC00086 that supported the development of SpIRIT in Australia and the currently ongoing Moon to Mars Initiative Demonstrator Mission Grant that is supporting the operations of SpIRIT. We would also like to thank Thales Cryogenics and DcubeD for the cryocooler and pin puller systems that feature in the TheMIS design. Titomic has played a pivotal role in the manufacturing of the copper cold-sprayed radiator, the authors would like to thank the local Victorian manufacturing facility and the manufacturing engineer that prepared the SpIRIT flight radiators: Khin Thar. Several interns of the Melbourne Space Laboratory team have worked on various elements of TheMIS; a special thanks to Fynn Oppermann, Jonathan Green, and Oliver Vogel-Reed.